# Numerical Simulation of Fibre Suspension Flow through an Axisymmetric Contraction and Expansion Passages by Brownian Configuration Field Method

Zhumin Lu, Boo Cheong Khoo*, Hua-Shu Dou, Nhan Phan-Thien, and Khoon Seng Yeo

Department of Mechanical Engineering
National University of Singapore
Singapore 119260, SINGAPORE

*Corresponding author: mpekbc@nus.edu.sg

**Abstract**

In this paper, the finite element method is combined with the Brownian Configuration Field (BFC) method to simulate the fibre suspension flow in axisymmetric contraction and expansion passages. In order to solve for the high stress at high concentration, the Discrete Adaptive Viscoelastic Stress Splitting (DAVSS) method is employed. For the axisymmetric contraction and expansion passages with different geometry ratios, the results obtained are compared to available constitutive models and experiments. The predicted vortex length for dilute suspensions agrees well with experimental data in literature. Our numerical results show clearly the effect on vortex enhancement with increase of the volume fractions and the aspect ratios. Effect of aspect ratio of fibres on the vortex length is also studied. It is found that for the lower expansion ratio flows the vortex dimension in the corner region is fairly independent of fibre concentration and aspect ratio of fibres while the said vortex dimension increases with the increase of fibre concentration for contraction flows. The finding suggests that the aligned fibre approximation traditionally employed in previous work does not exactly describe the effect of fibre motion, and the present BFC method is deemed more suitable for the flow of dilute fibre suspensions. In terms of numerics, the employment of DAVSS enhances numerical stability in the presence of high concentration of fibre in the flow.

## 1. Introduction

Most fibre-reinforced materials are usually produced through injection moulding of fibre suspension. These materials are widely used in aeronautical and various industrial purposes.

Therefore, it is very important to understand the flow effects on the fibre orientation, and consequently, their effects on the mechanical properties of the final composite components when the mould is cured.

Due to the demand for high quality production and coupled with reduced manufacturing costs, numerical method for flow analysis is widely used in the design and manufacture of fibre-reinforeced material. The reliability of the commercial software for such analysis is highly dependent on the constitutive modeling of the fibre suspension flows. To describe the rheological properties of fibre suspensions, a few constitutive models are established [1,2,3]. In these models, the contributions of fibre are treated as the ensemble-averaged behaviour via orientation tensors. Strictly, the motion of fibres in such flow is much more complex than those described. On the other hand, for a dilute suspension, the motion of fibre can be described reasonably by Jeffery's equation [4] because it is assumed that there is no interaction between the fibres. The fibre suspension is generally classified as dilute, semi-dilute and concentrated as follows. A suspension of uniform fibres can be characterized by its volume fraction $\phi$ and the fibre aspect ratio $a_r$ (the ratio of the fibre length $l$ to the fibre diameter $d$). If $\phi a_r^2 < 1$, the suspension is said to be dilute. For a dilute suspension, each fibre can move freely and there is no interaction between fibres. Hence the volume-averaged stress can be obtained by Jeffery's solution [4]. If $1 < \phi a_r^2 < a_r$ and $1 < \phi a_r$, the suspension is said to be semi-dilute and concentrated, respectively. Flogar and Tucker [5] proposed a model, in which the fibre interactions, modeled as a diffusion process, are added into the Jeffery's equation.

To simulate the flow of fibre suspensions, a common method is to solve the evolution equation of orientation tensor coupled with the flow field. However, some additional approximations are needed to close the evolution equation. The simplest closures are linear and quadratic approximations [6]. More complex closure approximations have also been proposed [7]. The main drawback of the closure approximation lies in its uncertainty. Phan-Thien and Graham [8] proposed another approximation based on the Geffery's equation. Recently, there have increasing works relating to the fibre suspension simulation [9-12], but most if not all of these employed the method in which the stress tensor is solved. In a separate development, based on the Brownian Configuration Field (BCF) method [13-14], Fan et al. [15] reported a robust numerical method to simulate flows of fibre suspensions. The BCF method does not need any closure approximation and can obtain good quality solutions at high volume concentrations of fibres. Fan et al. [16] also used the method to simulate the fibre suspension flows with shear-induced migration and the numerical result compared very well with experimental data. As emphasized before, the reliability of the numerical simulation depends to a large extent on the modelling of the constitutive behaviours. Therefore, it appears that the Brownian Configuration Field (BCF) method presents as a viable choice for this closure [17].

Contraction and expansion flows are common features in injection molding and other industrial processes. Fibre suspension induced vortex structure which occurs in both contraction and expansion flows has attracted much attention in the past [18-20]. There is still dearth of work on the flows of non-dilute fibre suspensions through even on axisymmetric contraction [19,20]. In this paper, we investigate the flow of fibre suspensions in a Newtonian solvent through an axisymmetric contraction and expansion using the finite element method combined with the BCF method, which has been successfully used by Fan et al. [15,16]. In the closing of the governing equations, two constitutive models due to Lipscomb et al. [18] and Phan-Thien and Graham [21] are employed, respectively. Vortex structure and formation in both expansion and contraction are investigated for fibre suspension flows. Effects of different contraction and expansion ratios, the volume fractions and the aspect ratio of fibres are studied.

## 2. Governing Equations and Preliminaries

The motion of incompressible continua is governed by mass and momentum conservation equation, written in dimensionless form as

$$\nabla \cdot \boldsymbol{u} = 0, \tag{1}$$

$$\frac{\partial \mathrm{u}}{\partial t} + \mathrm{u} \cdot \nabla \mathrm{u} = -\nabla p + \frac{1}{\mathrm{Re}} \nabla \cdot \tau \tag{2}$$

where $\boldsymbol{u}, t, p$ and $\boldsymbol{\tau}$ are the dimensionless velocity, time, pressure and extra-stress, respectively, are given as,

$$\mathrm{u} = \frac{\mathrm{u}}{\mathrm{U}}, \quad \mathrm{t} = \frac{\mathrm{tU}}{\mathrm{L}}, \quad \mathrm{p} = \frac{\mathrm{p}}{\rho \mathrm{U}^2}, \quad \text{and} \quad \tau = \frac{\tau}{\rho \mathrm{U}^2}.$$

Re is the Reynolds number defined as

$$\mathrm{Re} = \rho UL/\eta_0 ,$$

where $\rho$ is the density of fluid; $\eta_0$ is the viscosity of suspending fluid which is assumed to be a Newtonian fluid; $U$ and $L$ are the characteristic velocity and length scales, respectively.

The extra-stress tensor, $\tau$, can be decomposed as follows,

$$\boldsymbol{\tau} = \boldsymbol{\tau}_N + \boldsymbol{\tau}_f, \tag{3}$$

where $\boldsymbol{\tau}_N$ is the Newtonian contribution from the suspending fluid and $\boldsymbol{\tau}_f$ is the contribution from the fibres. For $\boldsymbol{\tau}_N$, it can be written as

$$\boldsymbol{\tau}_N = 2\boldsymbol{D}, \tag{4}$$

where $\boldsymbol{D} = (\nabla \boldsymbol{u} + \nabla \boldsymbol{u}^T)/2$ is the dimensionless rate-of-deformation tensor. There are a few constitutive equations to describe the fibre stress in Newtonian solvent. In this study, two models are adopted. One is the model for dilute suspensions developed by Lipscomb et al. [19], which is written as

$$\boldsymbol{\tau}_f = \frac{\phi \mu}{\eta_0} \boldsymbol{D} : \langle \boldsymbol{PPPP} \rangle, \tag{5a}$$

where $\phi$ is the volume fraction of fibres, $\mu$ is a material constant, $\boldsymbol{P}$ is a unit vector along the fibre's direction, $\langle \cdots \rangle$ denotes the ensemble average over the orientation space of fibres, and $\langle \boldsymbol{PPPP} \rangle$ is the fourth-order orientation, called structure tensor. Here $\phi a_r < 1$ valid for dilute suspension where $a_r$ is the aspect ratio of the fibre. Another is the modified transversely isotropic fluid (TIF) model by Phan-Thien and Graham[8], combining with Folgar-Tucker's model [5] to account for the random interaction between fibres. It should be pointed out that this model can be used for the case of non-dilute suspension flows [8] where $1 < \phi a_r^2 < a_r$ (semi-dilute) or $1 < \phi a_r$ (concentrated). This model has been applied to simulate complex flows of fibre suspensions including shear-induced migration [8, 15]. It can be expressed as

$$\boldsymbol{\tau}_f = f(\phi)\left[ A\boldsymbol{D} : \langle \boldsymbol{PPPP} \rangle + 2D_r F \langle \boldsymbol{PP} \rangle \right], \tag{5b}$$

with

$$f(\phi)=\frac{\phi(2-\phi/\phi_m)}{2(1-\phi/\phi_m)^2}, A=\frac{a_r^2}{\ln 2a_r-1.5}, F=\frac{3a_r^2}{\ln 2a_r-0.5},\tag{6}$$

where $D_r$ is the diffusion coefficient, $\langle PP\rangle$ are the second-order structure tensor, and $\phi_m$ is basically the maximum volume fraction of the suspension. The latter can be approximated by the linear relation [8]

$$\phi_m=0.53-0.013a_r, 5<a_r<30.\tag{7}$$

In this paper, the first constitutive model is applied to the flow of dilute suspensions in order to compare our results with those of Lipscomb et al. [19] and Chiba et al. [20].

To describe the evolution of fibre's orientation in non-dilute suspensions, Folgar and Tucker [5] added a diffusion term to the moment equation. This is equivalent to adding Brownian noise to the Jeffery's equation,

$$\frac{d\boldsymbol{P}_{(i)}}{dt}=L\cdot\boldsymbol{P}_{(i)}-L:\boldsymbol{P}_{(i)}\boldsymbol{P}_{(i)}\boldsymbol{P}_{(i)}+\left(\boldsymbol{I}-\boldsymbol{P}_{(i)}\boldsymbol{P}_{(i)}\right)\cdot\boldsymbol{F}^{(b)}(t),\qquad i=1,\ldots N\tag{8}$$

where $\boldsymbol{P}_{(i)}(i=1,2,\cdots N)$ is the unit vector along the axis of the $i$th fibre, $L$ is the effective velocity gradient tensor, $L\equiv\nabla\boldsymbol{u}^T-\zeta\boldsymbol{D}$, with $\zeta=2/(a_r^2+1)$ and the random force with properties

$$\langle\boldsymbol{F}^{(b)}(t)\rangle=0\tag{9}$$

and

$$\langle\boldsymbol{F}^{(b)}(t+s)\boldsymbol{F}^{(b)}(t)\rangle=2D_r\delta(s)\boldsymbol{I}.\tag{10}$$

In the above equation, $\delta(s)$ is the Dirac delta function and $\boldsymbol{I}$ is the unit tensor. Folgar and Tucker [5] make the following constitutive assumption:

$$D_r=C_i\dot{\gamma},\tag{11}$$

where $\dot{\gamma} = \sqrt{2(\boldsymbol{D}:\boldsymbol{D})}$ is the generalized strain rate, $C_i$ is the interaction coefficient which depends on the volume fraction of fibre suspensions and aspect ratio of fibres and may be treated as a second-order tensor[21]. For simplicity, we shall take $C_i$ as a constant.

The random force can be expressed in terms of the white noise,

$$F^{(b)}(t) = \sqrt{2C_i\dot{\gamma}}\frac{dw_t}{dt},$$

where $w_t$ is the Wiener process and it is a Gaussian random function.

In Eq. (8), $P_{(i)}$ is a random vector function of particle and time. Based on the idea of Brownian configuration field [13,14], it can be treated as a random vector field $P_{(i)}(x,t)$ which is a function of space and time. As mentioned in [13], a new configuration field $\boldsymbol{Q}_{(i)}(x,t)$ is introduced:

$$\boldsymbol{Q}_{(i)}(x,t) = Q_{(i)}\boldsymbol{P}_{(i)}(x,t), \tag{12}$$

where $Q_{(i)}$ is the modulus of $\boldsymbol{Q}_{(i)}(x,t)$. From Eq. (8), The evolution equation of $\boldsymbol{Q}_{(i)}(x,t)$ can be derived:

$$\frac{\partial \boldsymbol{Q}_{(i)}}{\partial t} + \boldsymbol{u}\cdot\nabla\boldsymbol{Q}_{(i)} = \boldsymbol{L}\cdot\boldsymbol{Q}_{(i)} + Q_{(i)}\boldsymbol{F}^{(b)}(t) \tag{13}$$

Expressing the random force with the white noise and integrating the above equation from $t$ to $t+\Delta t$, we obtain its Euler scheme with first-order weak convergence [22]:

$$\boldsymbol{Q}_{(i)}(x,t+\Delta t) + \boldsymbol{u}\cdot\nabla\boldsymbol{Q}_{(i)}(x,t+\Delta t)\Delta t = \boldsymbol{Q}_{(i)}(x,t) + \boldsymbol{L}\cdot\boldsymbol{Q}_{(i)}(x,t)\Delta t + \sqrt{2C_i\dot{\gamma}}\boldsymbol{Q}_{(i)}(x,t)\Delta w(t)$$

(14)

where $\Delta w(t)$ is the increment of the Wiener process. The structure tensor can be calculated approximately by the ensemble average:

$$\langle \boldsymbol{P} \cdots \boldsymbol{P} \rangle = \frac{1}{N} \sum_{i=1}^{N_p} \left\langle \frac{\boldsymbol{Q}_{(i)}}{Q_{(i)}} \cdots \frac{\boldsymbol{Q}_{(i)}}{Q_{(i)}} \right\rangle. \tag{15}$$

## 3. Numerical Method

The Galerkin finite element method is employed to solve for the governing equations of fluid flows. The contribution of fibre motion to the stress tensor is accounted for by the Brownian Configuration Field (BFC) method. In order to solve for the high stress at high concentration, the Discrete Adaptive Viscoelastic Stress Splitting (DAVSS) method [23] is employed. It was shown that this method developed is a very stable and robust even when the stress is high [23,15,25]. According to the idea of the DAVSS, the momentum equations can be re-written as

$$\frac{\partial \mathrm{u}}{\partial t} + \mathrm{u} \cdot \nabla \mathrm{u} - \nabla \cdot \left[ \eta_a \left( \nabla \mathrm{u} + \nabla \mathrm{u}^T \right) \right] = -\nabla p - \nabla \cdot \left[ (\eta_a - 1) \mathrm{G} \right] + \frac{1}{\mathrm{Re}} \nabla \cdot \boldsymbol{\tau}_f, \tag{16}$$

where $\boldsymbol{G}$ is the strain rate tensor, which is a new unknown:

$$\boldsymbol{G} = \nabla \boldsymbol{u} + \nabla \boldsymbol{u}^T. \tag{17}$$

and $\eta_a$ is the adaptive viscosity. The selection of $\eta_a$ will following the work of Sun et al. [23] and Fan et al. [15] giving rise to

$$\eta_a = \frac{\phi \mu}{\eta_0} + \frac{1 + \sqrt{(1/2) \boldsymbol{\tau}_f : \boldsymbol{\tau}_f}}{1 + \sqrt{(1/2) \boldsymbol{G} : \boldsymbol{G}}}, \tag{18}$$

for the fibre stress model of Lipscomb et al. [19]. For the model of Phan-Thien and Graham [8], $\eta_a$ is expressed as

$$\eta_a = A f(\phi) + \frac{1 + \sqrt{(1/2) \boldsymbol{\tau}_f : \boldsymbol{\tau}_f}}{1 + \sqrt{(1/2) \boldsymbol{G} : \boldsymbol{G}}}. \tag{19}$$

Let $\Omega$ denote the flow domain and $\partial\Omega$ its boundary. The flow domain is discretized by a set of non-overlapping quadrilateral elements. Let $V^h$, $P^h$ and $G^h$ represent the discretized function spaces over $\Omega$. Taking $V^h(\Omega) \in Q_2(\Omega)$, $P^h(\Omega) \in Q_1(\Omega)$ and $G^h(\Omega) \in Q_1(\Omega)$ where $Q_2(\Omega)$ and $Q_1(\Omega)$ are the spaces of biquadratic and bilinear polynomials, the Galerkin weak formulation for mass and momentum equations is given as follows: Find $\left(u^{h(n+1)}, p^{h(n+1)}, G^{h(n+1)}\right) \in V^h \times P^h \times G^h$ at $t = t_{n+1}$ such that for all $(v, q, g) \in V^h \times P^h \times G^h$,

$$\left(v, \mathrm{Re}\left(\frac{u^{h(n+1)}}{\Delta t}\right)\right)_\Omega + \left(\nabla v, \eta_a \left(\nabla u^{h(n+1)} + \left(\nabla u^{h(n+1)}\right)^T\right)\right)_\Omega - \left(\nabla \cdot v, p^{h(n+1)}\right)_\Omega$$
$$= (v, t)_\Omega + \left(v, \mathrm{Re}\left(\frac{u^{h(n)}}{\Delta t} - u^{h(n)} \nabla u^{h(n)}\right)\right)_\Omega + \left(\nabla v, (\eta_a - 1) G^{h(n)}\right)_\Omega + \left(\nabla v, \tau_f\right)_\Omega \quad (20)$$

$$\left(q, \nabla \cdot u^{h(n+1)}\right)_\Omega = 0 \quad (21)$$

$$\left(g, G^{h(n+1)} - \nabla u^{h(n+1)} - \left(\nabla u^{h(n+1)}\right)^T\right)_\Omega = 0. \quad (22)$$

Here v, q and g are the weighting functions; $t$ is the traction at the boundary of the flow domain; $(.,.)$ denotes the inner product; superscript $n$ and $n+1$ denote the time step index. The nonlinear inertial terms are represented explicitly because the Reynolds number is very small in our simulations.

The Brownian configuration field (Eq.(14)) is discretized using the discontinuous Galerkin(DG) method, which can be stated as follows: Find $Q_{(i)}^{h(n+1)} \in S^h$ such that for all $s \in S^h$, for $i = 1, \ldots, N_f$

$$\left(s, Q_{(i)}^{h(n+1)} + u_{(i)}^{h(n+1)} \nabla Q_{(i)}^{h(n+1)} \Delta t\right)_{\Omega_e}$$
$$= \left(s, Q_{(i)}^{h(n)} + L \cdot Q_{(i)}^{h(n)} \Delta t + \sqrt{2 C_i \dot{\gamma}} Q_{(i)}^{h(n)} \Delta w\right)_{\Omega_e} - \left(s, n \cdot u_{(i)}^{h(n)} \cdot \left(Q_{(i)}^{+h(n)} - Q_{(i)}^{h(n)}\right) \Delta t\right)_{\partial\Omega_e^-} \quad (23)$$

where $\Omega_e$ is the domain of element, $\partial\Omega_e^-$ is the inflow boundary of this element with the normal vector $n$ and $N_f$ is the number of configuration field. $Q_{(i)}^h$ is the finite dimension approximation of random vector field $\boldsymbol{Q}_{(i)}$ and $\boldsymbol{Q}_{(i)}^{+h}$ is its value in the upstream neighboring element and $S^h(\Omega) \in Q_1(\Omega)$.

The initial flow is assumed to be Newtonian fluid and the initial configuration field is assumed to be random. After solving Eq.(23) element by element, the fibre stress and the adaptive viscosity at $n+1$ time step can be obtained, and then the flow field at $n+1$ time step is solved from Eqs.(20)-(23).

**4. Problem Definitions**

The abrupt axisymmetric contraction geometry is described in Fig. 1. The fluid flows from a large circular tube through an abrupt entry into a smaller circular tube. The contraction ratio is defined as

$$\beta = \frac{R_U}{R_D}, \qquad (24)$$

where $R_U$ and $R_D$ are the radii of upstream and downstream tubes respectively. The dimensionless re-attachment length is defined as

$$\chi = \frac{L_v}{2R_U}, \qquad (25)$$

where $L_v$ is the re-attachment length of upstream vortex as shown in Fig. 1. We use $R_U$ as the characteristic length and the mean velocity in the upstream tube as the characteristic velocity. Contrary to the contraction flows, in expansion flows the expansion ratio is defined as $\beta = R_D/R_U$ and $\chi = L_v/2R_U$. In expansion flows, $R_D$ is chosen to the characteristic length and the mean velocity in the downstream tube as the characteristic velocity.

A major interest in contraction flows is the total pressure loss which has high implication in injection molding process. Let $\delta P$ denote the pressure difference between the entry and exit section, $\Delta P_U$ and $\Delta P_D$ the upstream and downstream fully developed pressure gradients, respectively, $L_U$ and $L_D$ the lengths of the upstream and downstream tube and $\tau_w$ the fully developed, wall-shear-stress in the downstream tube. The Couette correction $C$ is then defined as follows which is used to characterize the feature of the total pressure loss:

$$C = \frac{\delta P - \Delta P_U L_U - \Delta P_D L_D}{2\tau_w} \qquad (26).$$

In our calculations, $L_U = 6$ and $L_D = 5$ for the contraction flows while $L_U = 6$ and $L_D = 4$ for the expansion flows. Three meshes (M1, M2, M3) shown in Fig. 2 are employed in the simulations. On mesh M1 the total number of elements, nodes and degrees of freedom are (600, 2521, 7240). For mesh M2 and M3, these are (1200, 4961, 14452) and (1925, 7911, 23170), respectively. The full developed velocity profiles are imposed in entry and exit sections. No-slip boundary conditions are imposed on the solid wall and axisymmetric boundary condition is applicable on the axis of symmetry. The computations are first carried out for the straight section with Reynolds number 0.01 and the time step is set at 0.001. The number of Brownian configuration fields $N_f$ is set to 1000. Some of the predicted Couette correction with time for the three meshes is shown in Fig. 3. Figure 3(a) shows the predictions using the stress model of Lipscomb and the parameter $\phi\mu/\eta_0$ is chosen as 2, 5 and 8 respectively. The results using the stress model of Phan-Thien and Graham with two groups of parameters are presented in Figure 3(b). Both Fig. 3(a) and Fig. 3(b) show that Couette correction on all three meshes converge and the fluctuations decrease with mesh refinement as described in [15]. From these figures, it can be seen that the flow is approximately steady for $t > 1$. In the subsequent computations, mesh M3 is used to obtain the solutions and the results at $t = 6$ are collected. It is also shown in Fig. 3 that the fluctuations in the Couette correction become stronger with the increasing aspect ratios and volume fractions. This is because the statistical errors in the fibre stress increase due to the limited number of fibres for large stresses. In the case of high aspect ratios and high volume fractions, it is necessary to increase the number of Brownian configuration fields [15].

**5. Numerical Results with Model due to Lipscomb et al.**

There are simulations of the axisymmetric contraction flows of dilute fibre suspensions [18-20]. In all of these, the model developed by Lipscomb et al.[18] was adopted to describe the properties of fibre suspension. In the simulations, the quadratic closure approximation is usually used relative to the fourth-order tensor $\langle PPPP \rangle$ and the second-order tensor $\langle PP \rangle$:

$$\mathbf{D}:\langle PPPP \rangle = \mathbf{D}:\langle PP \rangle\langle PP \rangle. \qquad (27)$$

Furthermore, it is assumed that the fibres are fully aligned with the flow:

$$\langle PP \rangle = uu/|u \cdot u|, \qquad (28)$$

where $uu$ is the dyadic product of velocity vector $u$. In the finite element solution, creeping flow of fibre suspension through an axisymmetric 4:1 contraction was considered [18]; a limiting value at $\phi\mu/\eta_0 = 11.3$ was noted with mesh refinement (the solution did not converge beyond this). Chiba et al. [19] employed the ADI method and SOR method to solve the vorticity and streamfunction equations, respectively, and they also encountered a computation limiting value of $\phi\mu/\eta_0 = 12$.

To compare with the predictions of Lipscomb et al. [18] and Chiba et al. [19], we employ the BCF method to get $< PPPP >$ in the simulation of axisymmetric contraction flow of dilute fibre suspensions using the stress model of Lipscomb et al.[18] (Eq.(5a)). We do not encounter the numerical limiting value of $\phi\mu/\eta_0$ but there is a need to increase the number of Brownian configuration field with increasing $\phi\mu/\eta_0$.

Compared with the numerical prediction and experimental measure of Lipscomb et al.[18], the computed dimensionless vortex length by BCF method for 4.5:1 contraction flow, shown in Fig. 4 for $\phi\mu/\eta_0 \leq 5$, is in good agreement with experiments. For $\phi\mu/\eta_0 > 5$, the predicted vortex length is lower than that of the experiments; this may be due to the fact that the experiment conditions are outside the applicable range of the constitutive model [18]. As seen in Fig. 4, our results are better than Lipscomb's computations referring to the experimental data. The may suggest that the aligned fibre approximation employed in the literature [18] does not exactly describe the effect of fibre motion, and the BFC method is more suitable for the flow of dilute fibre suspensions. In Fig. 5, a comparison of the present results using BCF method with the predictions of Chiba et al. for 4:1 contraction flow is shown. The present result is in excellent agreement with Chiba et al.'s for $\phi\mu/\eta_0 \leq 3$, and two results diverges for $\phi\mu/\eta_0 > 3$. A possible reason for the disagreement between the sets of numerical data at high values of $\phi\mu/\eta_0$ may lie in the aligned fiber approximation utilized in both Lipscomb et al.'s simulation [18] and Chiba et al.'s [19], as discussed above for Fig.4.

The contours of the streamline for the 4:1 contraction are shown in Fig. 6 for $\phi\mu/\eta_0 = 2$, $\phi\mu/\eta_0 = 5$ and $\phi\mu/\eta_0 = 8$. With increasing $\phi\mu/\eta_0$ of the solution, the shape of vortex boundary becomes more convex to the main flow; this is consistent with the results of in literature [18, 19]. However, the experimental results [18,19] showed that the vortex boundary (meaning the separating line between the vortex and the main flow) is almost a straight line for all the flow rate.

## 6. Numerical Results with Model due to Phan-Thien and Graham

### 6.1 Results for Contraction Flow

In this section, we will discuss the numerical results of a 4:1 axisymmetry contraction using the stress model of Phan-Thien and Graham [21] for non-dilute fibre suspensions. These simulations are carried out using mesh M3. Figure 7 and Figure 8 show the streamline contours for $a_r = 10$ and $a_r = 20$, respectively. The recirculating vortex obtained is exhibited in the streamline contours. In these figures, the vortex boundary is nearly straight for all the volume fraction of fibres. This behaviour is agreement with that found from the experiments [18,19]. The vortex structures simulated are qualitatively similar to the flow visualization observation of Lipscomb et al. [18].

In the following, the numerical results for contraction ratio $\beta$ =2, 4 and 8 will be compared and discussed. Figure 9 show the streamline contours for $a_r = 10$ and $\phi = 0.15$ for ratio $\beta$ =2, 4 and 8 and Figure 10 show those for $a_r = 20$, $\phi = 0.10$. From these figures, the dimensionless vortex length for $\beta = 4$ is larger than those of $\beta$ =2 and 8 for the same flow parameters. The boundary of vortex zone depicts a rather straight section for $\beta$ =2 and 4 but displays a more convex shape for $\beta = 8$ in Fig 9 and Fig 10. This phenomenon may be attributed to the influence of the pressure drop across the contraction. At high contraction ration, the acceleration in the region near the centerline leads to a large pressure drop along the axis. This pressure drop near the axis around the position of geometry contraction forms a zone of low pressure. On the other hand the pressure along the wall is not much affected by the acceleration of flow near the axis and the thus the pressure in the corner (A in Fig.1) is not changed much by the contraction ratio as in position B. Therefore, for high contraction ratio ($\beta$ =**8**), a big pressure difference is generated between point B and point A, and the strong back flow is driven by this pressure difference.

Figure 11 shows the dependence of the dimensionless vortex length on the volume fractions for two fibre aspect ratios of $a_r = 10$ and $a_r = 20$ under three different contraction ratios. In general, the vortex length increases with the increasing of $\phi$. At a given $\phi$, the vortex length increases with the increase of the aspect ratio of fibre. However, for $\beta$=2, the aspect ratio $a_r$ has definite influence on the vortex length at $\phi$<0.15. At $\phi$>0.15, the influencing of $a_r$ is greatly diminished. This phenomenon needs further studying.

To examine the total pressure loss through the axisymmetric contraction under different contraction ratio, variation of Couette correction with the volume fraction is shown in Fig. 12. For all the contraction ratios, the pressure loss increases with the increase of the volume fraction and the fibre aspect ratio; these trends are also observed by other researchers for contraction flows of dilute suspension [18]. As with the vortex length, the Couette correction also tends towards an asymptotic constant with increasing volume fractions. The reason is that the term $f(\phi)$ in the stress model becomes very large when $\phi \to \phi_m$ and there is relatively weak effect on stress from the fibre aspect ratio. It is very interesting to note that for the contraction flow of the same suspension, the pressure loss for $\beta$=4 is lower than those for $\beta$=2 and 8, and the corresponding vortex length of $\beta$=4 is the largest**.**

### 6.2 Results for Expansion Flow

In contrast to the contraction flows, the fibre suspension flows through an abrupt expansion have received relatively little attention in the literature. In [25], Townsend and Walter reported some experimental and numerical results for dilute suspensions. In this section, the expansion flows for different expansion ratios will be presented. These simulations are carried out using mesh M3.

Figures 13 and 14 show the streamline contours for $a_r = 10$, $\phi = 0.15$ and $a_r = 20$, $\phi = 0.10$, respectively, for the expansion flows. It can be seen that the vortex dimension increases with increasing $\beta$ for the same parameters of suspensions. This tendency can be attributed to the pressure increase across the expansion due to the expansion effect. It is also noticed that the vortex dimension in expansion flow is less than that of contraction flow for the same ratio. The predicted streamlines in expansion flows has been compared with the experimental observations of [25] for some flow and geometrical parameters. For dilute fibre suspension, qualitative agreement is obtained for both the contraction and expansion flow. Figure 15 shows the dependence of the dimensionless vortex length on the volume fractions for two fibre aspect ratio of $a_r = 10$ and $a_r = 20$ for three different expansion ratios. For $\beta = 2$ and 4, there is almost no change of the vortex length as the volume fraction increases and the effect of $a_r$ is very small. On the other hand, the vortex length increases for $\beta = 8$ and the effect of $a_r$ is also imminent. This dramatic change in the tendency of vortex length versus the fibre fraction needs further studies. An intuitive relation is that for high expansion flows, the variation of fibre fraction or aspect ratio leads to large variation of stresses as well as the pressure. The dependence of pressure on the fibre parameters results in the alteration of the vortex length for high expansion ratio.

Figures 16 shows the effect of the number of Brownian configuration fields on the accuracy of simulations for the expansion flow with geometry of $\beta = 2$. The mesh used for these calculations is coarse M1. The values of the normal stress $\tau_{f_{11}}$, the normal stress $\tau_{f_{22}}$, the shear stress $\tau_{f_{12}}$ and the first normal stress difference $N_1 = \tau_{f_{11}} - \tau_{f_{22}}$ are recorded at the position E (see Fig.1, for the three numbers of Brownian configuration fields at $N_f$ =500, 1000, and 2000, as shown in Figs. 16(a), (b), (c), and (d), respectively. All of these figures clearly depict that there is a starting period before the flows assume the quasi-steady state. For the mesh employed, the minimum number of time steps for this period is about 2500. After this period, the stresses vary in the vicinity of an averaged value as time progresses. It is observed from Fig.16(a) that there is a limited oscillation for $N_f$ =500 while the amplitude of oscillation has increased for the normal stress $\tau_{f_{22}}$ as can be seen in Fig.16(b). On the other hand, the counterpart for $N_f$ =1000 and $N_f$ =2000 have depicted corresponding lower level of oscillation. These observations suggest that increasing the number of Brownian configuration fields enhances the accuracy of the computation, although it may be mentioned that there is no marked or significant difference between $N_f$ =1000 and $N_f$ =2000. As such, it can be accepted that most of the calculations carried out in this work are for $N_f$ =1000 without loss of fidelity. For the shear stress $\tau_{f_{12}}$ shown in Fig.16(c), the effect of the value of $N_f$ on the variation in the quasi-steady state is rather limited and quite indistinguishable from each other. Finally, for the first normal stress difference $N_1$ which is the difference between Fig.16(a) and Fig.16(b) and depicted in Fig.16(d), its value at $N_f$ =500 indicates relatively the largest variation while $N_f$ =1000 and $N_f$ =2000 show much smaller variation and almost indistinguishable from each other.

**7. Conclusions**

In this paper, the vortex structure and formation are investigated by FEM/BCF methods for fibre suspension flows through both expansion and contraction passages. With the constitutive model of Lipscomb et al.[18], the predicted vortex length using the BCF method is in good agreement with the experiment data of Lipscomb et al. [19] for $\beta = 4.5$ at $\phi\mu/\eta_0 \leq 5$, which is better than the numerical predictions of Lipscomb et al. [19] and Chiba et al. [20]. This may suggests that the aligned fibre approximation employed in [19, 20] does not exactly describe the effect of fibre motion. The present study using BCF method can reasonably capture the behaviour of fibre suspension.

With the constitutive model of Phan-Thien and Graham, the simulation results using the BCF method for flows through contraction and expansion flows are compared and discussed for different geometric and flow parameters (contraction and expansion ratios, the volume fractions and the aspect ratio). The flow structures in these two flows are quite different. For contraction flows, as a whole, the vortex length increases with the increasing of $\phi$. At a given fraction of fibre, the vortex length increases with the increase of aspect ratio of fibre. For expansion flows, the vortex dimension does not increase with the increase of the volume fraction and the aspect ratio for expansion flow at the lower expansion ratio ($\beta$=2 and 4). This observation is quite different from the vortex development in contraction flows in which the vortex dimension increases with the increase of the volume fraction and the aspect ratio for all contraction ratio ($\beta$=2, 4 and 8). The simulation results also show that the total pressure loss increase with the increase of the volume fractions and the fibre aspect ratios. At high contraction ratio or expansion ratio, the variation of pressure field generated by the fibre stresses may produce some changes of flow behaviour. These results may provide useful information for the design of moulds for use in injection molding of fibre suspension flows.

**Acknowledgements**
The authors thank the helpful discussions with XJ Fan at The University of Sydney and R Zheng at Moldflow company in Melbourne. The authors also wish to thank the financial support from A*STAR and Moldflow company for the project (Project No. EMT/00/011).

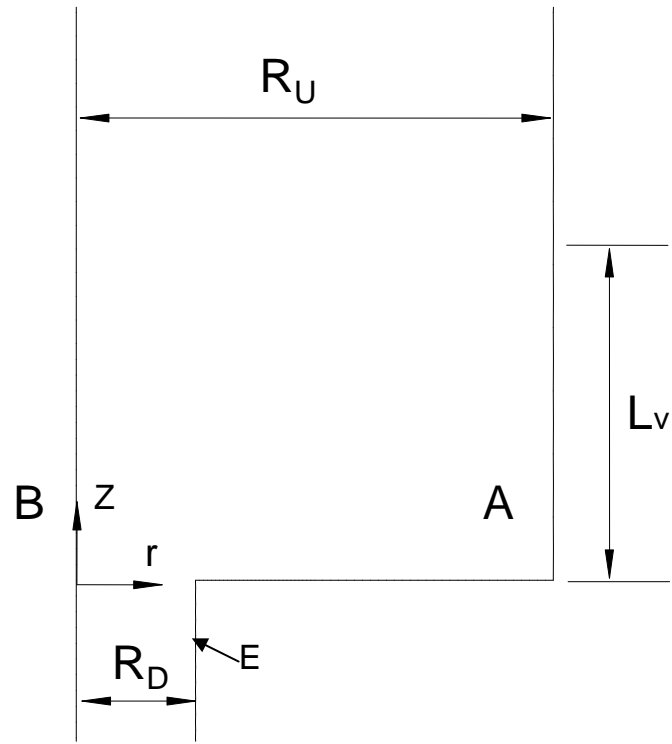

Fig. 1. Axisymmetric contraction flow geometry. The coordinate system is shown.

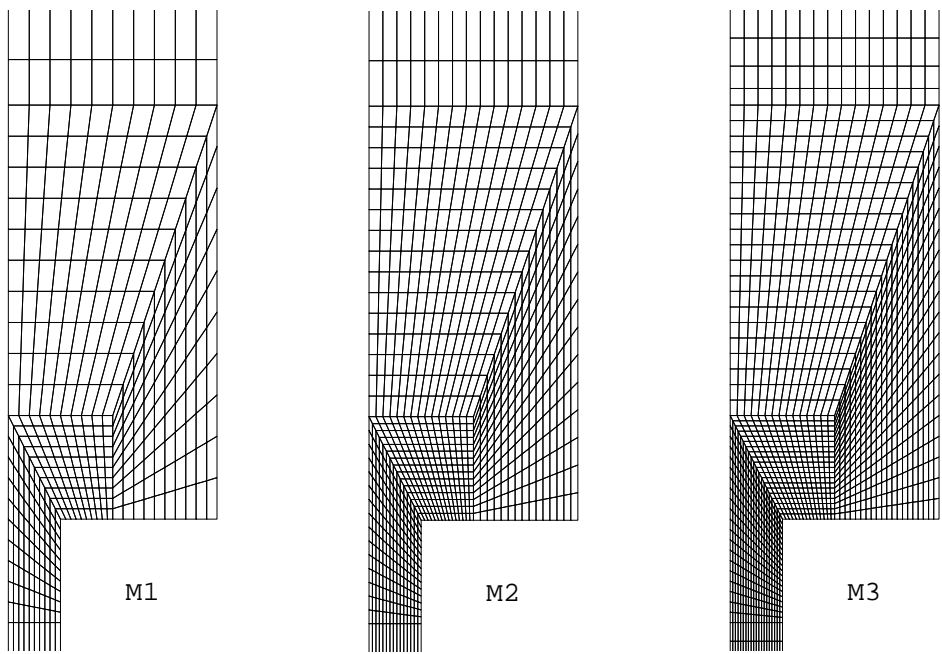

Fig. 2. The central portion of the three finite element meshes.

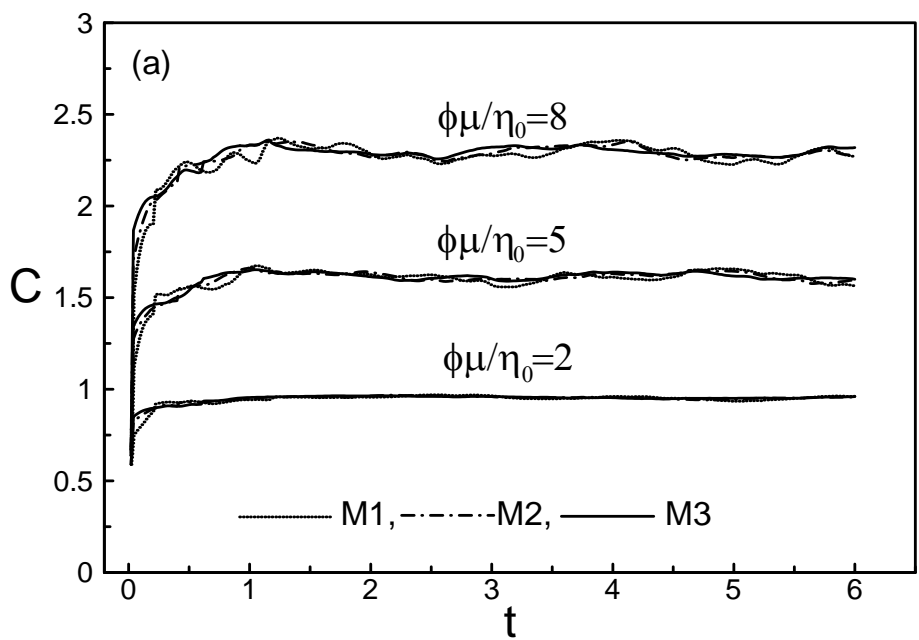

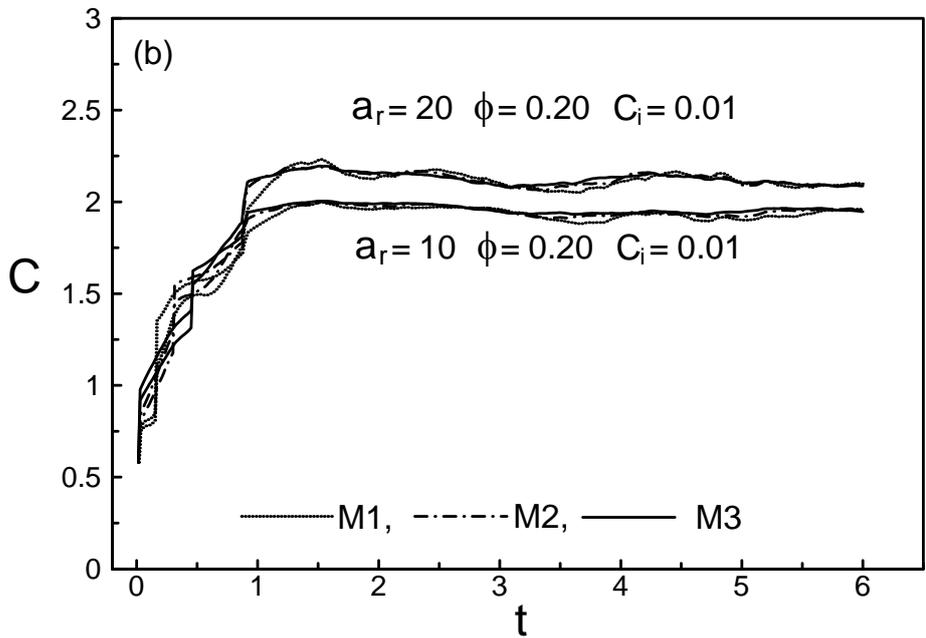

Fig. 3. Time evolution of Couette correction calculated using three finite element meshes, (a) for stress model of Lipscomb et al.; (b) for stress model of Phan-Thien and Graham

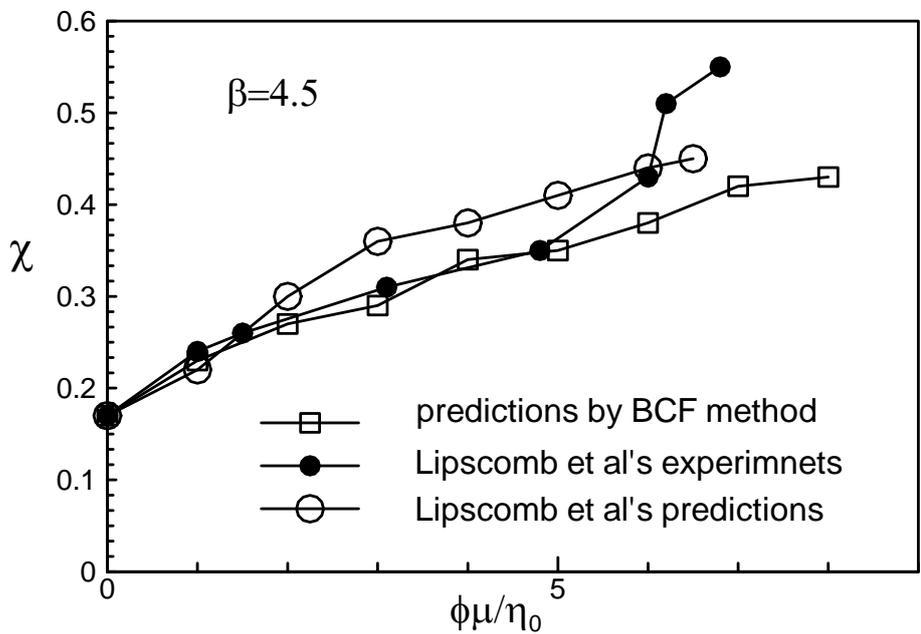

Fig. 4. Comparison between Lipscomb et al.'s vortex length and ours. $\beta = 4.5$

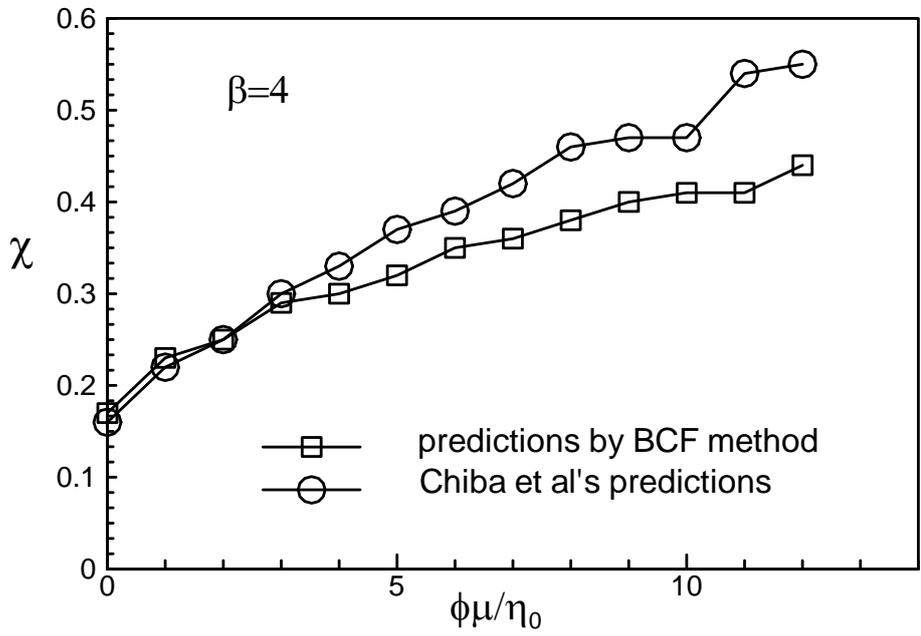

Fig. 5. Comparison of vortex length between Chiba et al.'s predictions and ours, $\beta = 4$.

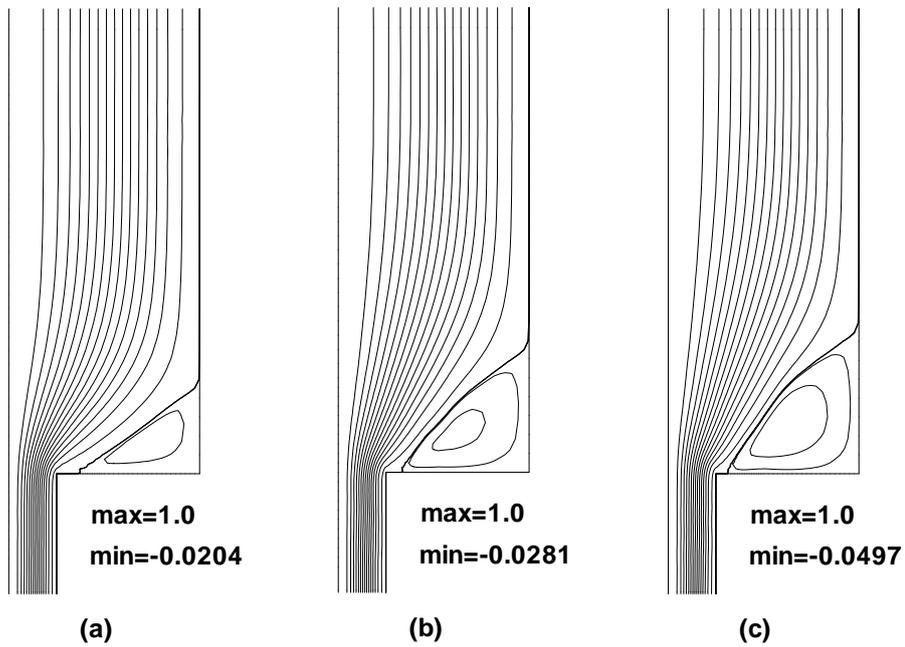

Fig. 6. Streamline contours for a 4:1 contraction. (a) $\phi\mu/\eta_0 = 2$; (b) $\phi\mu/\eta_0 = 5$; (c) $\phi\mu/\eta_0 = 8$.

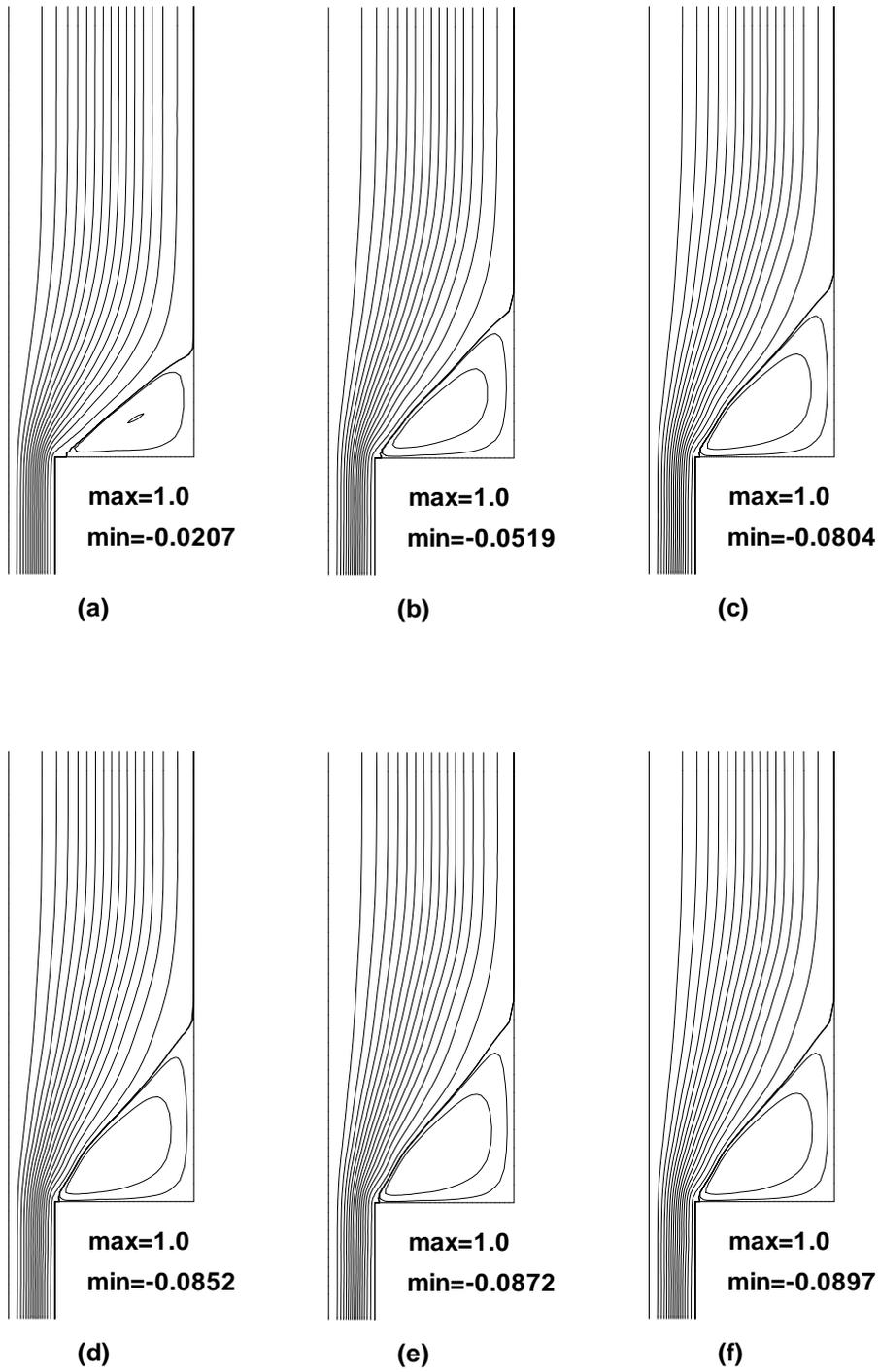

Fig. 7. Streamline contours with different $\phi$ for $a_r = 10$: (a) $\phi = 0.05$ ;(b) $\phi = 0.15$ ;(c) $\phi = 0.25$ ;(d) $\phi = 0.30$ ;(e) $\phi = 0.35$ ;(f) $\phi = 0.38$.

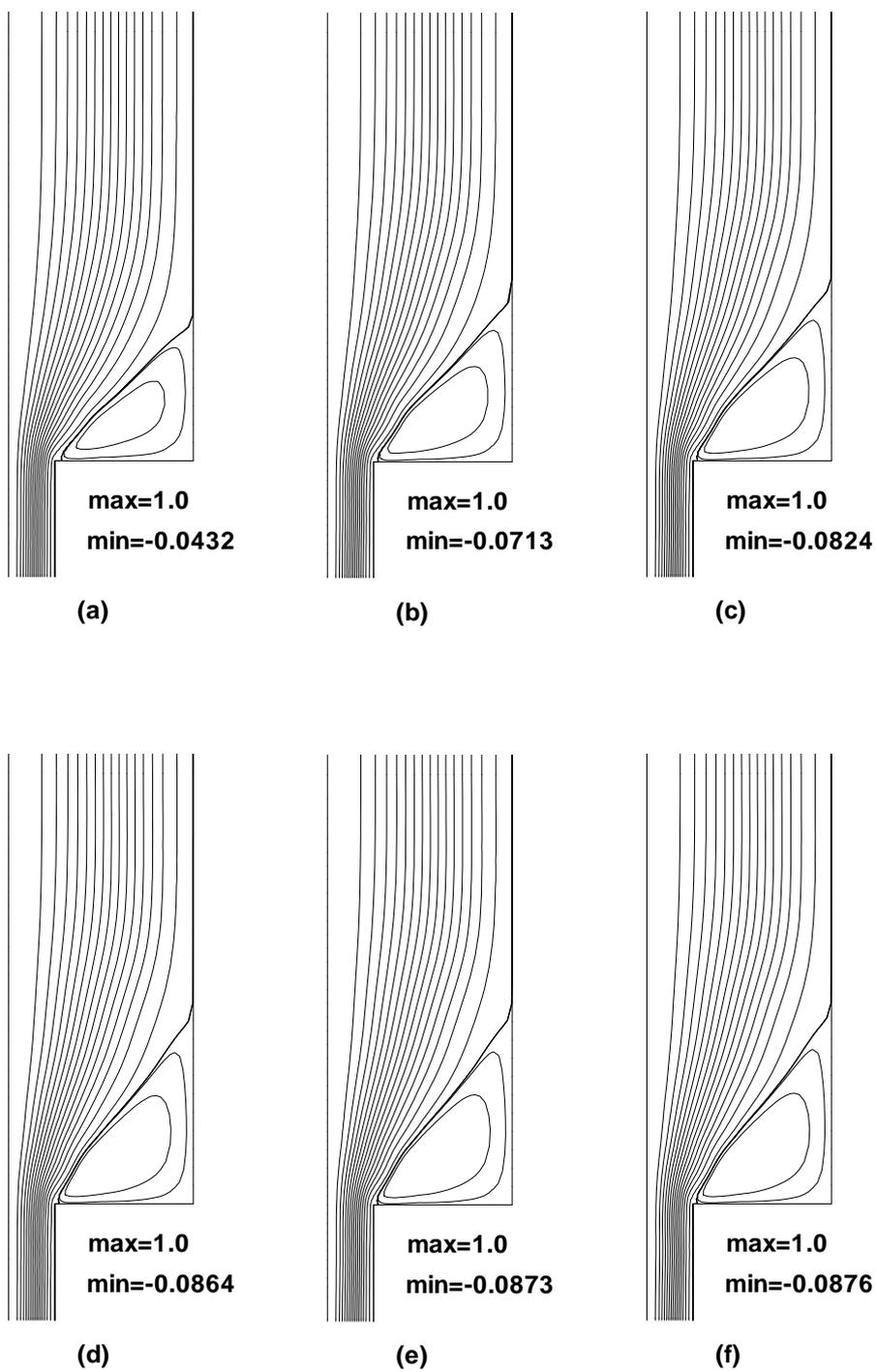

Fig. 8. Streamline contours with different $\phi$ for $a_r = 20$: (a) $\phi = 0.05$ ;(b) $\phi = 0.10$ ;(c) $\phi = 0.15$ ;(d) $\phi = 0.20$ ;(e) $\phi = 0.22$ ;(f) $\phi = 0.25$.

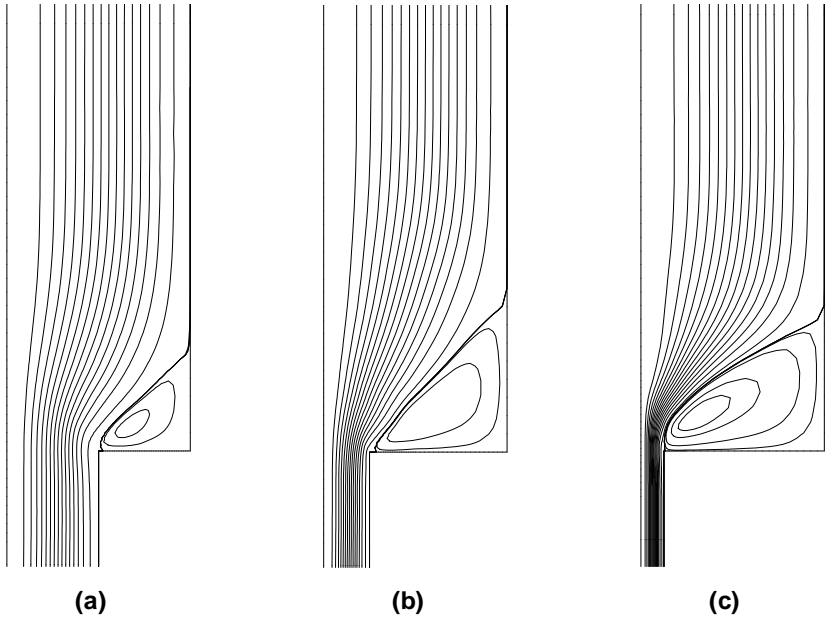

Fig 9. The streamline contours of contraction flows for $a_r = 10$ and $\phi = 0.15$. (a) $\beta = 2$; (b) $\beta = 4$; (c) $\beta = 8$.

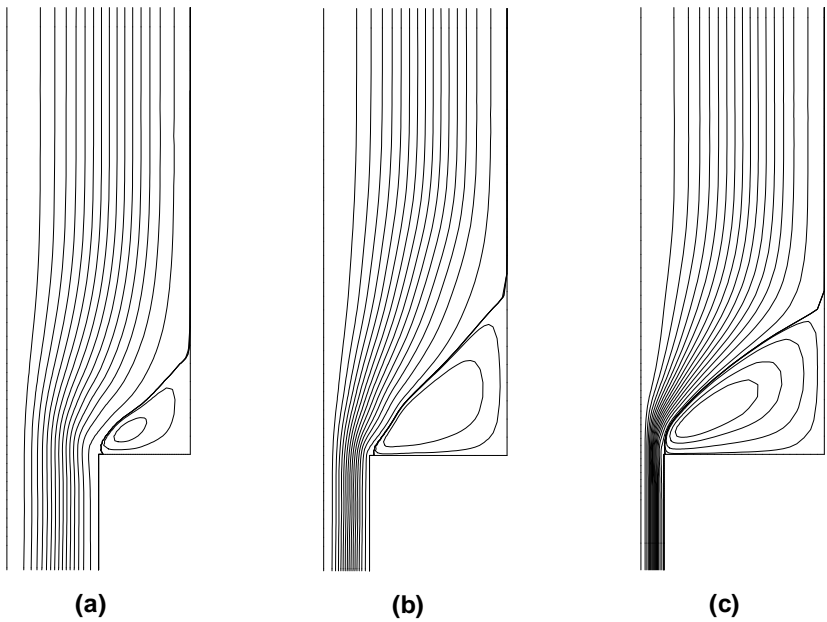

Fig.10. The streamline contours of contraction flows for $a_r = 20$ and $\phi = 0.10$. (a) $\beta = 2$; (b) $\beta = 4$; (c) $\beta = 8$.

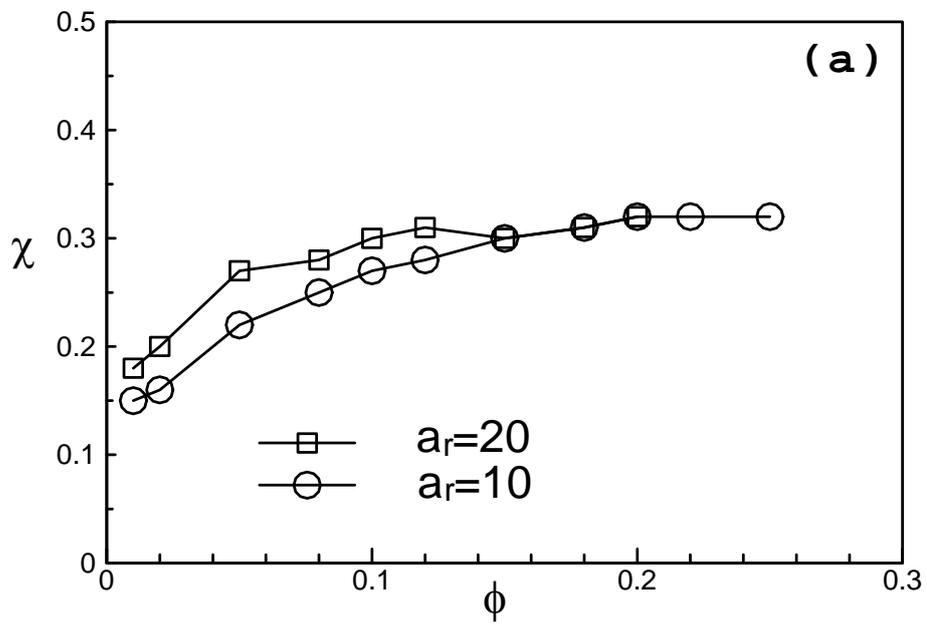

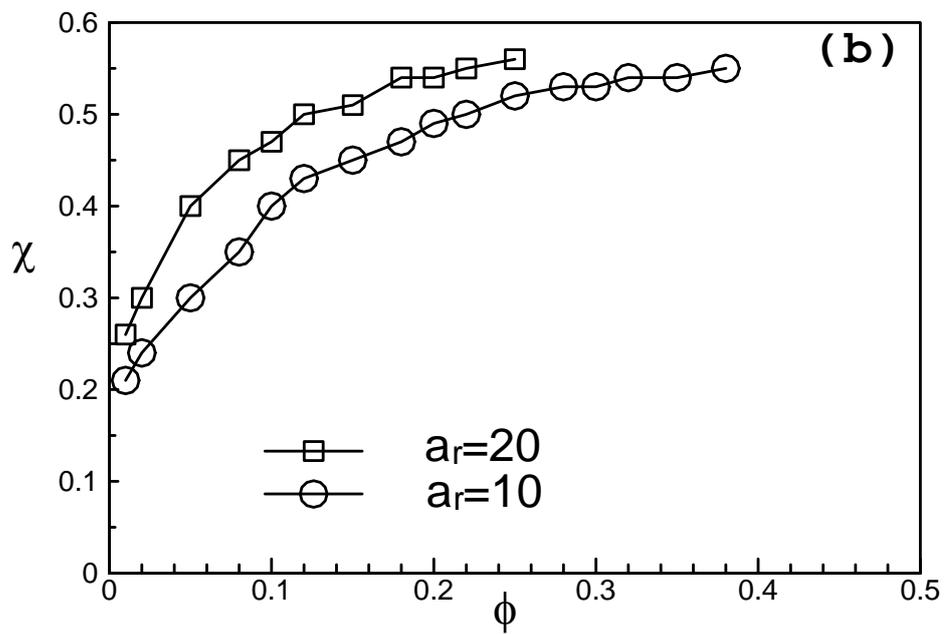

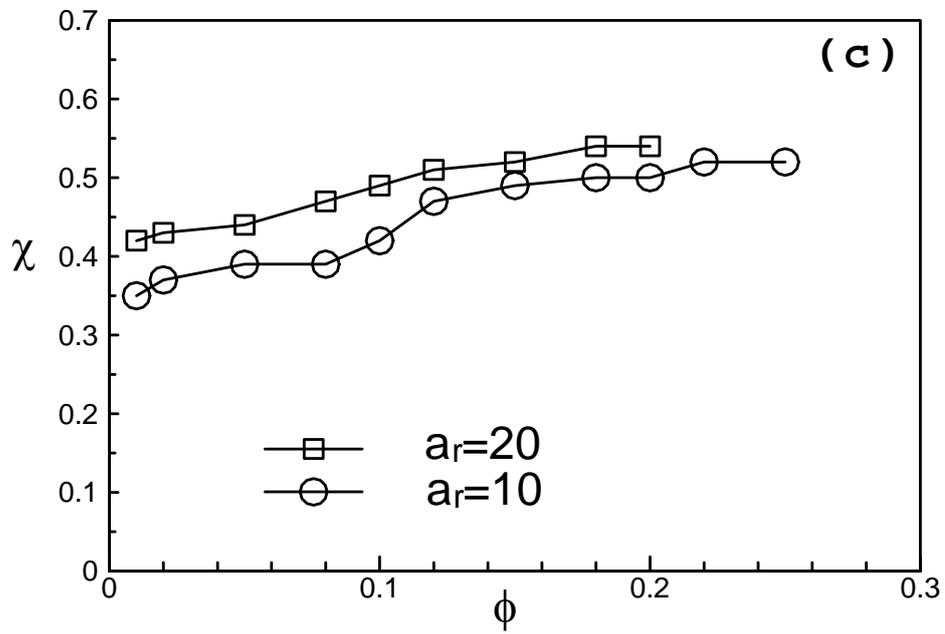

Fig. 11. Dependence of vortex length on $\phi$ for contraction flow. (a) $\beta = 2$; (b) $\beta = 4$; (c) $\beta = 8$.

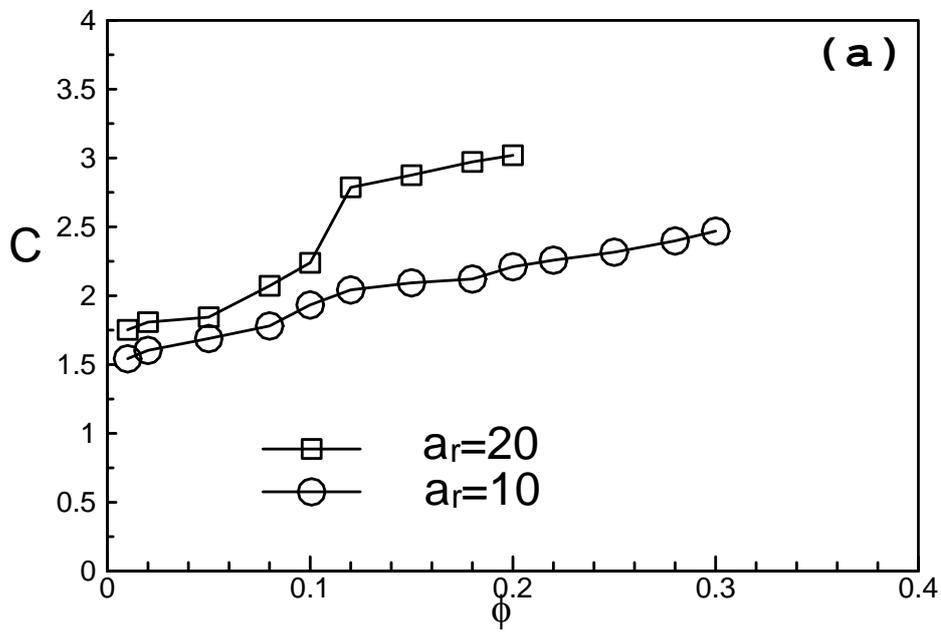

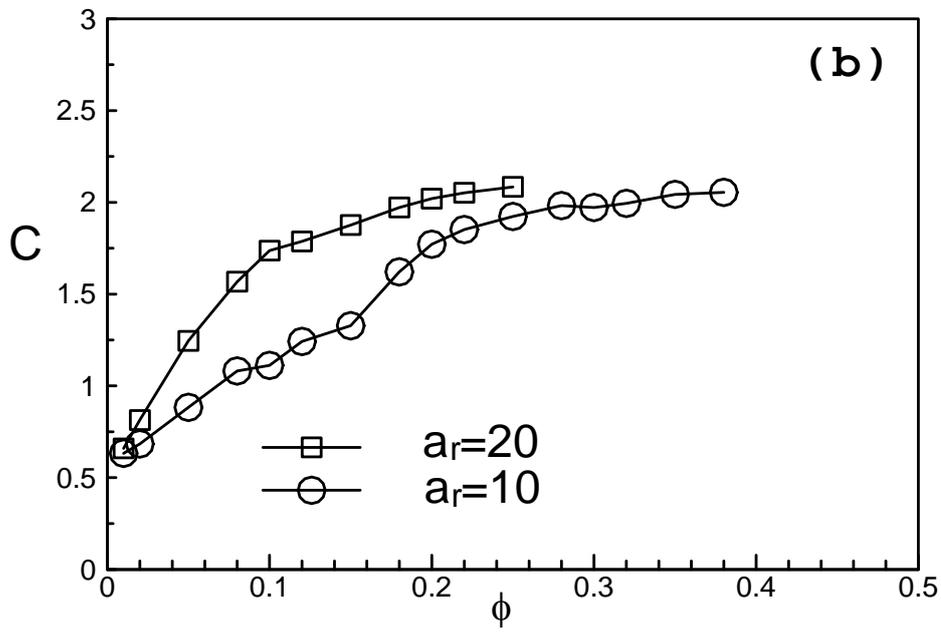

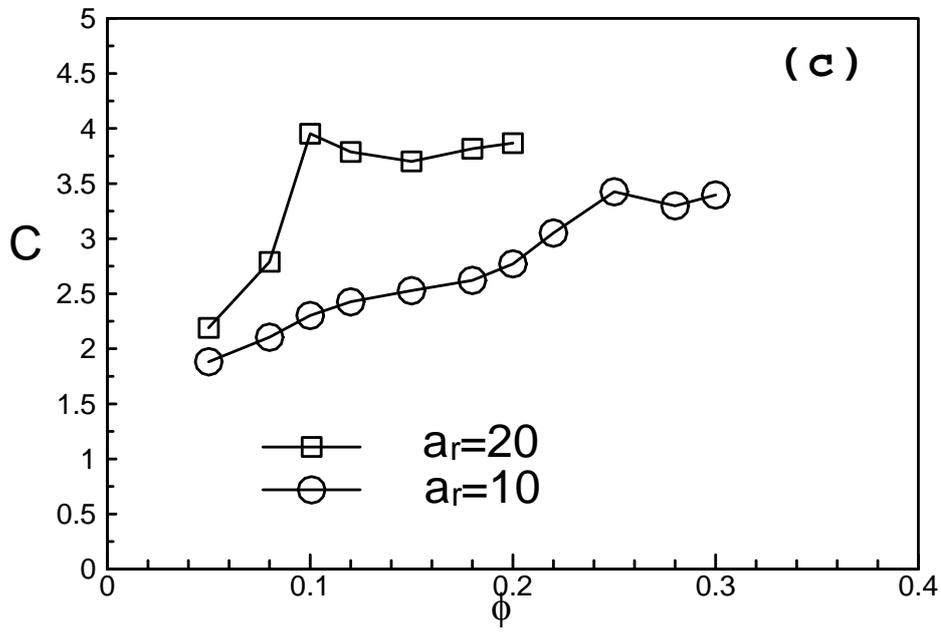

Fig. 12. Dependence of Couette correction on $\phi$ for contraction flows. (a) $\beta = 2$; (b) $\beta = 4$; (c) $\beta = 8$.

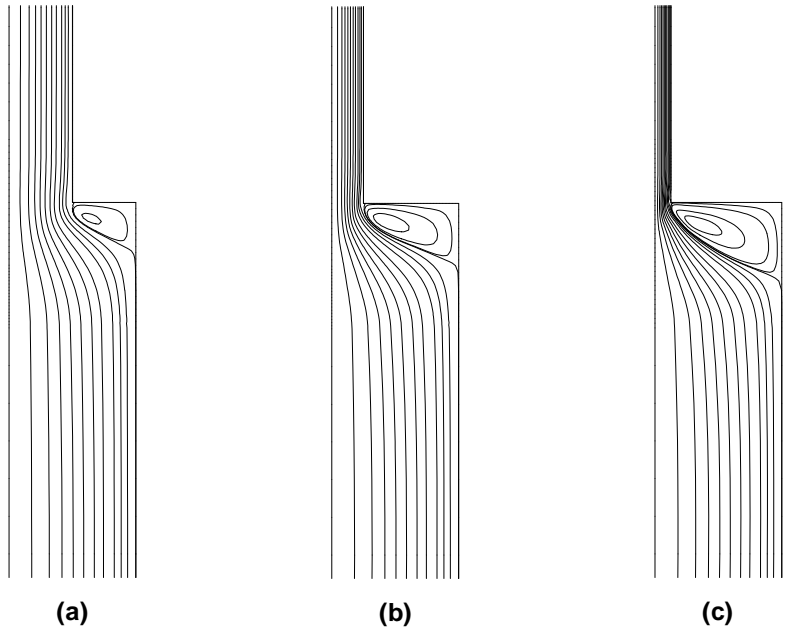

Fig 13. The streamline contours of expansion flows for $a_r = 10$ and $\phi = 0.15$. (a) $\beta = 2$; (b) $\beta = 4$; (c) $\beta = 8$.

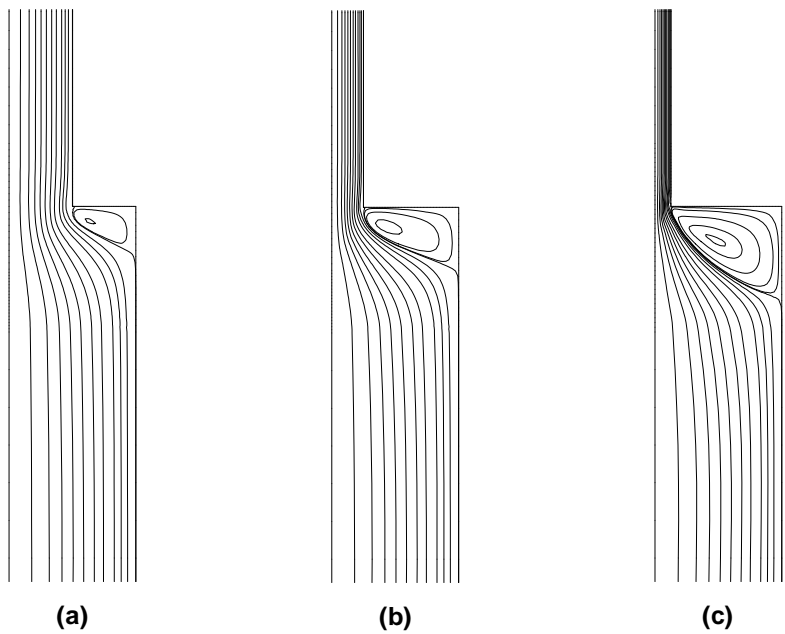

Fig.14. The streamline contours of expansion flows for $a_r = 20$ and $\phi = 0.10$. (a) $\beta = 2$; (b) $\beta = 4$; (c) $\beta = 8$.

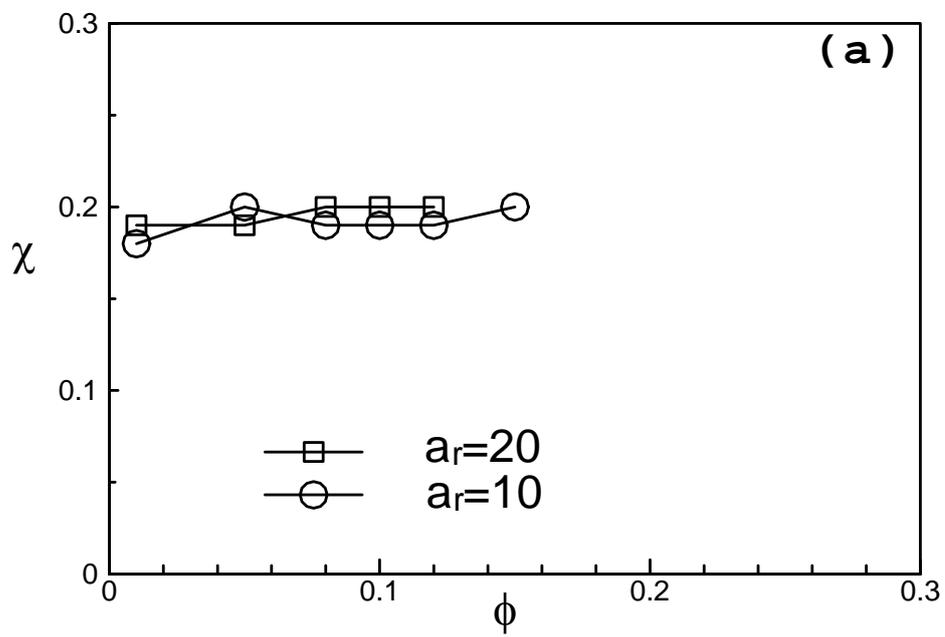
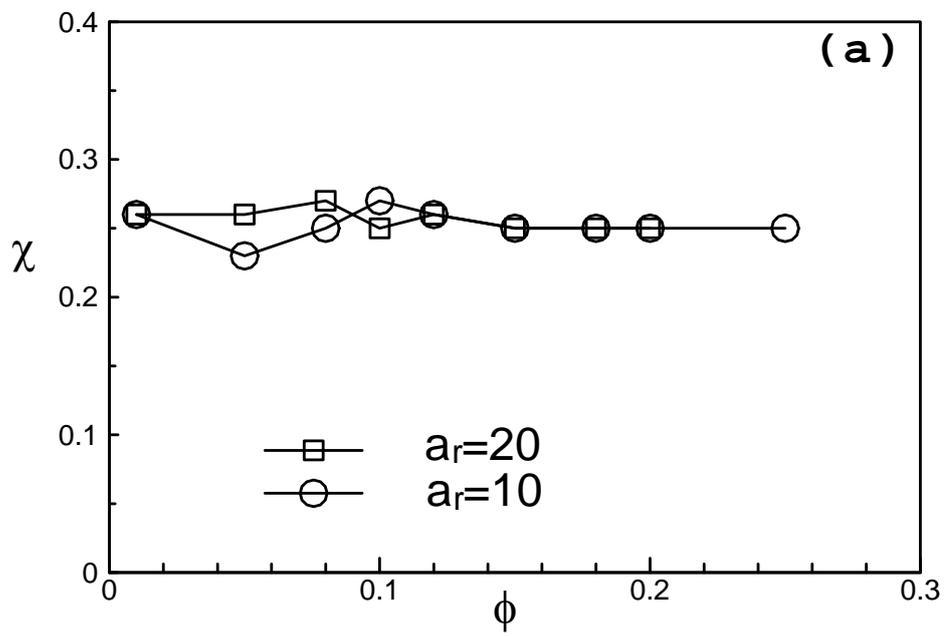

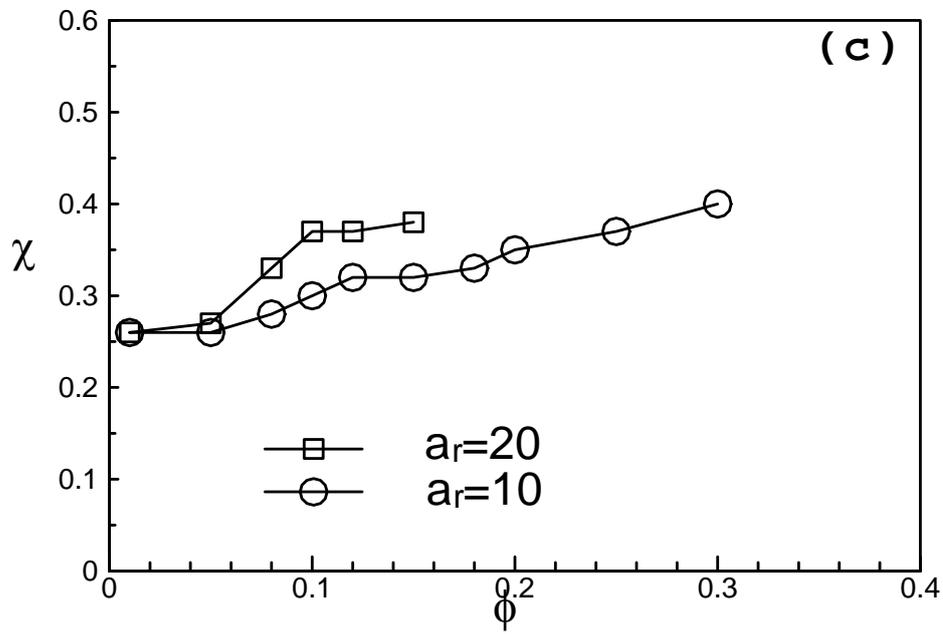

Fig. 15. Dependence of vortex length on $\phi$ for expansion flows. (a) $\beta = 2$; (b) $\beta = 4$; (c) $\beta = 8$.

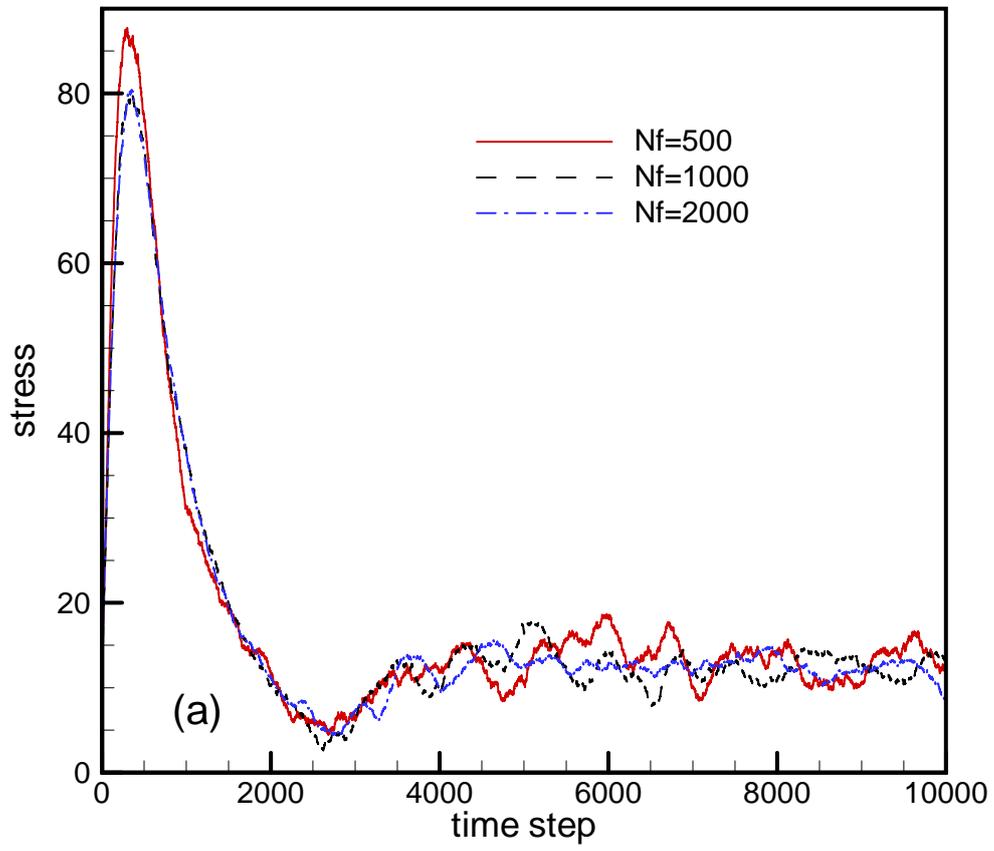

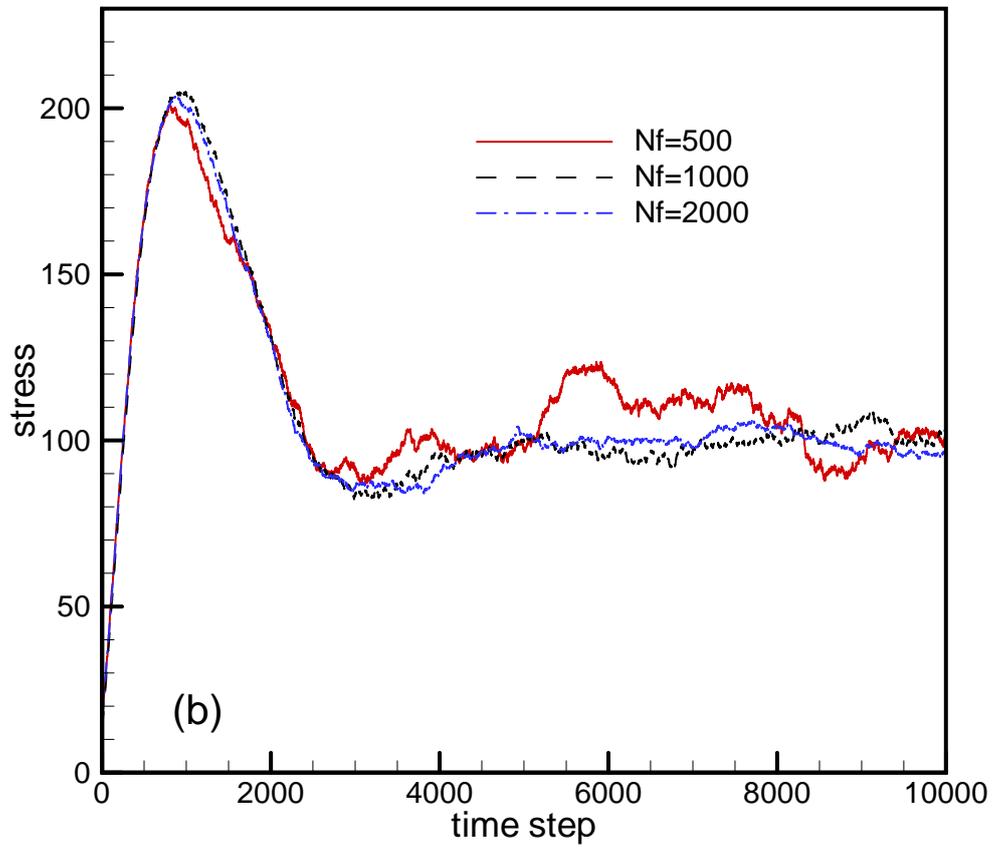

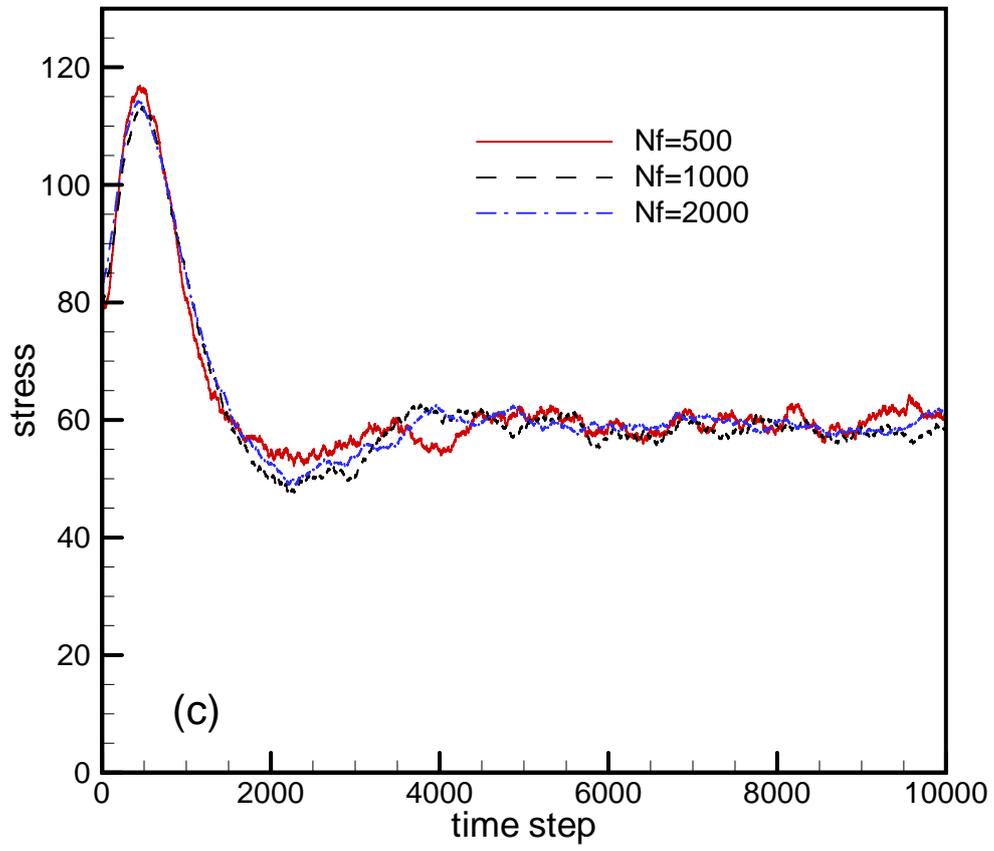

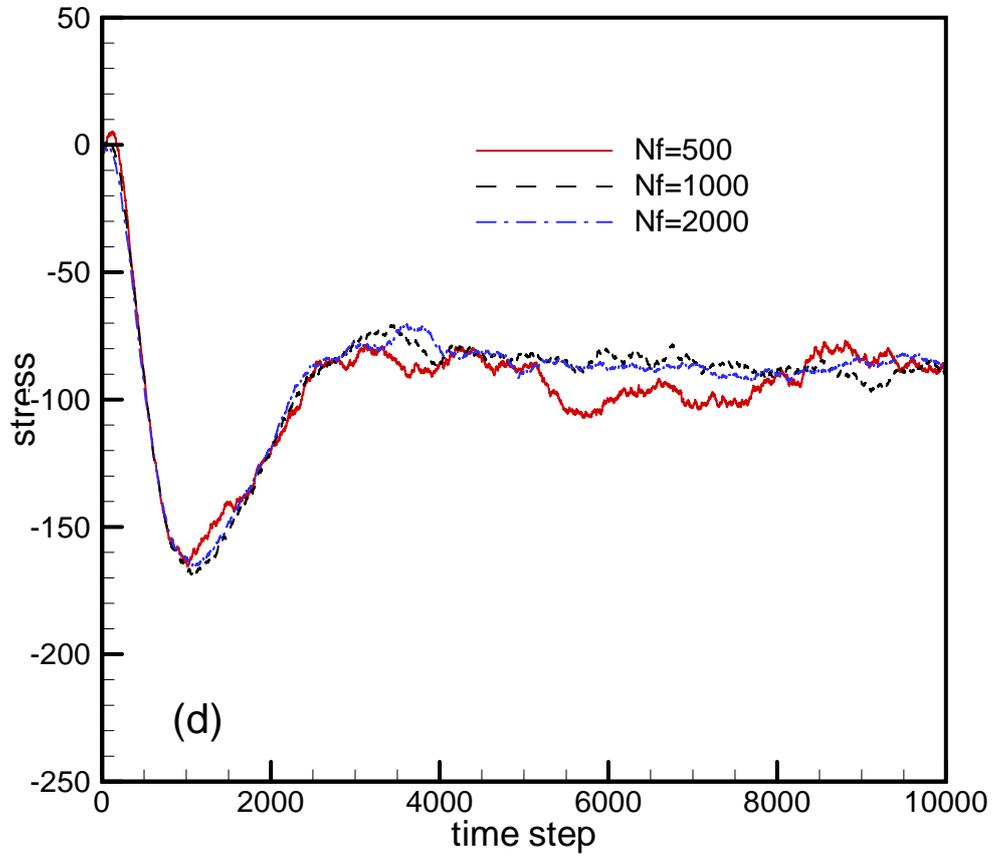

Fig. 16. Effect of the configuration number on the calculations for the expansion flow with geometry of $\beta = 2$. (a) The normal stress $\tau_{f_{11}}$; (b) The normal stress $\tau_{f_{22}}$; (c) The shear stress $\tau_{f_{12}}$; (d) The first normal stress difference $N_1 = \tau_{f_{11}} - \tau_{f_{22}}$. The values of these stresses are recorded at the position E as shown in Fig.1.